\documentclass [12pt,preprint]{aastex}

\usepackage{epsfig}
\begin{document}
\voffset-1cm
\newcommand{\gsim}{\hbox{\rlap{$^>$}$_\sim$}}
\newcommand{\lsim}{\hbox{\rlap{$^<$}$_\sim$}}

\title{The Discovery of a Hyperluminal Source in the Radio Afterglow of GRB 030329}

\author{Shlomo Dado\altaffilmark{1}, Arnon Dar\altaffilmark{1} and
A. De R\'ujula\altaffilmark{2}}

\altaffiltext{1}{dado@phep3.technion.ac.il, arnon@physics.technion.ac.il, 
dar@cern.ch.\\
Physics Department and Space Research Institute, Technion, Haifa 32000, 
Israel}
\altaffiltext{2}{alvaro.derujula@cern.ch.
CERN, CH-1211 Geneva 23, Switzerland;
Physics Dept. Boston Univ., USA}


\begin{abstract}

Taylor, Frail, Berger and Kulkarni have made precise VLBI measurements
of the size and position of the source of the radio afterglow of GRB 030329.
They report a size evolution compatible with standard fireball models,
proper motion limits inconsistent with the cannonball model, and 
a double source, i.e.~``an
additional compact component'' on day 51 after the GRB, totally
unexpected in the standard models. We outline a consistent interpretation
of the ensemble of the data in the realm of the cannonball model.
The observed double source is a radio image of the two cannonballs
required in this model to explain the $\gamma$-ray and optical light
curves of this GRB; their separation agrees with the expectation.
Thus interpreted, the observation of the two sources ---separated by
a ``hyperluminal'' distance--- is a major discovery in astrophysics:
it pins down the origin of GRBs.

\end{abstract}


\section{Motivation}

The currently best-studied theories of Gamma-Ray Bursts (GRBs) and their
afterglows (AGs) are the {\it Fireball} models (see, e.g., Zhang \&
Meszaros 2003 for a recent review) and the {\it Cannonball} (CB) model
(see, e.g., Dar \& De R\'ujula 2003a; Dado, Dar \& De R\'ujula, 2002;
2003a and references therein). The first set of models is often considered
to be {\it the standard model} of GRBs. The Standard and CB models are 
different in their original basic hypotheses, in their description of 
the data, and in their predictions.  

The CB model (Dar \& De R\'ujula 2000, 2003a) is based on the
assumption that GRBs and their afterglow (AG) are produced by superluminal
CBs (plasmoids made of ordinary matter) ejected in the explosions of 
supernovae (SNe).
The predicted {\it bulk motion} of the CBs of GRBs is so fast ---in comparison to 
that of quasar or microquasar ejecta--- that we have dubbed it 
{\it ``hyperluminal"} (Dado, Dar \& De R\'ujula, 2003a). Superluminal motions 
of moving astronomical objects in the plane of the sky
were first discussed ---in exquisite detail and with 19th-century elegance--- 
by Courdec (1939). Following the discovery of the very bright 
GRB 030329, at a nearby $z=0.1685$, we previously studied the 
radio-detectability of the hyperluminal
motion of its ejecta (Dado, Dar \& De R\'ujula 2004, hereafter DDD). 

Taylor, Frail, Berger and Kulkarni (Taylor et al.~2004,
hereafter TFBK) have recently published their VLBI observations
of the radio AG of GRB 030329. They report that {\it ``In
observations taken 51 days after the burst we detect an additional
compact component at a distance from the main component of 0.28 $\pm$
0.05 mas (0.80 pc). The presence of this 
component is not expected from the standard model''}. TFBK argue 
that the CB model is also inconsistent with their observations.
We propose an interpretation of the data in which that is
not the case. The gist of our interpretation is simple:
as we argued in DDD,
there are {\it two} sources relevant to the radio AG: the 
{\it two} CBs responsible for the observed
{\it two-pulse} $\gamma$-ray fluence and the observed
{\it two-shoulder} optical AG. The superluminal  
motions of the two sources have apparent displacements that
differ in magnitude (DDD).
 TFBK's double radio source (whose significance is larger than
20$\sigma$) is seen when the two CBs have
similar fluences at 15.3 GHz. At other times and frequencies
the two contributions have not been clearly resolved.
This is not so surprising: the R-band non-SN contribution
to the AG (i.e.~one of the CBs) has been seen to rebrighten at
a date coincident with the double-source observation
(Matheson et al.~2003).
Compact objects associated with a core-collapse supernova 
have been resolved once before, in SN1987A
(Nisenson \& Papaliolios 1999). In our interpretation
these objects were also CBs emitted in opposite directions.
Neither of them was pointing accurately enough
towards the Earth for its induced GRB to be observable.

\section{Summary of the relevant data}

TFBK have localized the radio AG of GRB 030329, with a precision of 0.2 mas,
 at various radio frequencies, and at
times corresponding to approximately 3, 8, 25, 51 and 83 days after burst.
At two epochs, TFBK report an observed angular size of the source: 
$0.065\pm 0.022$ mas at 15.3 GHz and $0.077\pm 0.036$ at 22.2 GHz
on day 25, and $0.172\pm 0.043$ mas at 8.4 GHz on day 87.
The rest of the data yield only upper limits for this observable. 
These results are compatible with expectations
of standard models.

More intriguingly, TFBK report that {\it ``The Gaussian fits made 
to the measured visibility data as a function
of baseline length returned residuals consistent with noise in all but
one case. The fit to the 15 GHz observation on May 19} [day 51]
{\it  produces a
significant residual ($>$20 sigma) which is $\sim$30\% of the peak
flux density and offset to the northeast at}
\begin{equation}
\Delta\alpha_{1,2}=0.28 \pm 0.05\;{\rm mas}
\label{obs2}
\end{equation}
{\it from the main component. The exact nature of this second component
is not known but it would require an average velocity of 19$c$ to
reach its offset from the flux centroid.''} 

Finally TFBK solve for proper motion, obtaining 
a limit on the angular displacement of {\it an assumed
single source} during the duration of the observations:
\begin{equation}
\alpha[80\,{\rm days}] = 0.10\pm 0.14\;{\rm mas}< 0.3\;{\rm mas}.
\label{obs3}
\end{equation}
 It is relevant to our
 interpretation of the data that TFBK assumed a single source (\S 4 \& \S 5).


\section{A CB-model interpretation of the double-source separation}

We identify TFBK's 20$\sigma$ evidence for a
double source as a radio image of the two CBs of 
GRB 030329. At the time of this observation, the
{\bf ``faster'' CB} has moved $0.28 \pm 0.05$ mas 
away from the {\bf ``slower" CB}, the motion of the latter
being unobservably small during the whole duration
of the campaign. We argue that this
straightforward interpretation of the double source is
compatible with the rest of the observations.

\begin{figure}[]
\hskip 2truecm
\vspace*{0.8cm}
\vspace*{-.4cm}
\hspace*{-1.cm}
\epsfig{file=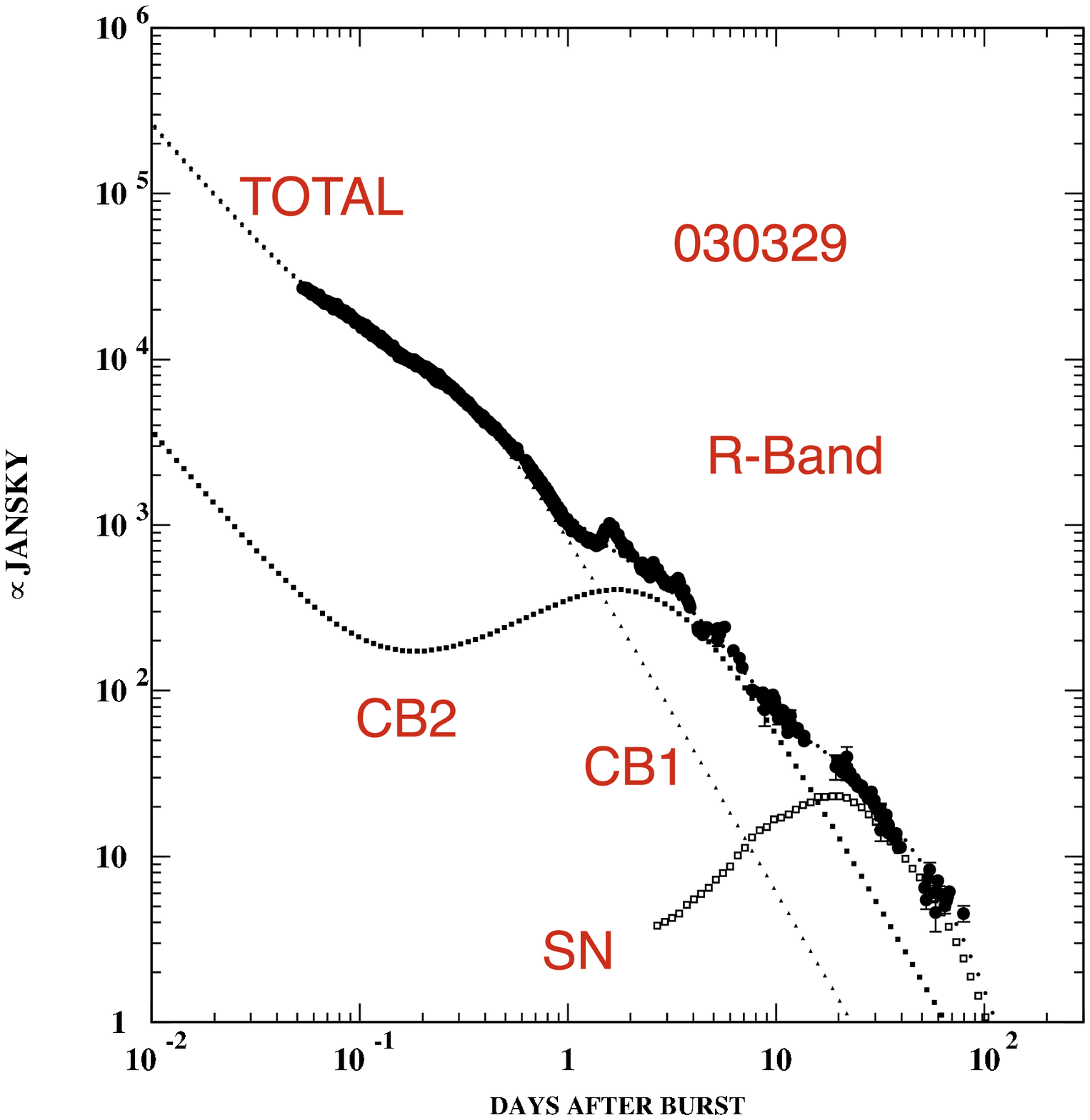, width=13.3cm}
\figcaption{The R-Band AG of GRB 030329 fit in the CB model with a
constant-density ISM and two CBs (DDD and references therein).
\label{figone}}
\end{figure}

As shown in Fig.~(\ref{figone}), the R-band
 optical AG of GRB 030329 was overtaken by the contribution
of the associated SN at day $\sim 10$ after the explosion
(Garnavich et al.~2003; Stanek et al.~2003), as expected
(Dado, Dar \& De R\'ujula, 2003b). 
The CB-model interpretation of this GRB (and, previously, that of
GRB 021004; Dado, Dar \& De R\'ujula, 2003c) requires
contributions from two CBs, producing
a two-pulse $\gamma$-ray fluence
and a two-shoulder optical AG. The predicted 
motion of the faster CB (labelled CB2 in the figure) is 
larger than that of the slower CB, CB1 (DDD). 

To a good
approximation, and for observer's times $t$ larger than a few
days, the approximate time dependence of a CB's angular displacement 
and velocity are:
\begin{eqnarray} 
\alpha_{_{\rm CB}}(t)&\approx& 
{\theta \over D_A}\;
\left[{6\;x_\infty^2\;c\,t \over (1+z)}\right]^{1/3}
\label{vcbang}\\
\dot\alpha_{_{\rm CB}}(t)&\approx&
{\theta \over 3\,D_A}\;
\left[{6\;x_\infty^2\;c \over t^2\,(1+z)}\right]^{1/3},
\label{omega} 
\end{eqnarray} 
where $D_A$ is the angular distance, $\theta$ is the angle 
of the CB's motion relative to the line of sight, and $x_\infty$
is a ``deceleration parameter" (Dado, Dar \& De R\'ujula, 2002). For an 
interstellar medium of constant baryon density, 
the above expressions are exact large-$t$ limits. 

The optical light curves of GRB 030329 are very finely structured,
as can be seen in Fig.~(\ref{figone}) for the case
of the R-band data.  In the CB model the sharp magnitude variations
starting at various times after day 1.5  were interpreted
and modelled in DDD as the result of density inhomogeneities that
the faster CB encounters as it exits the superbubble in which its parent
SN and many previous ones were immersed. The fits in DDD returned 
$\theta[1]\approx 2.2$ mrad and $\theta[2]\approx 2.3$ mrad
for the angles of the two CBs.
As the CBs cross the density inhomogeneities, the description of their
deceleration is quite elaborate (DDD), but the overall results for their
positions and velocities are still sufficiently well described by 
Eqs.~(\ref{vcbang},\ref{omega}), with ``effective'' parameters
$x_\infty[1]\approx 0.017$ Mpc and $x_\infty[2]\approx 0.048$ Mpc.

\begin{figure}[t]  
\begin{tabular}{cc}  
\hspace {1.7cm}
\epsfig{file=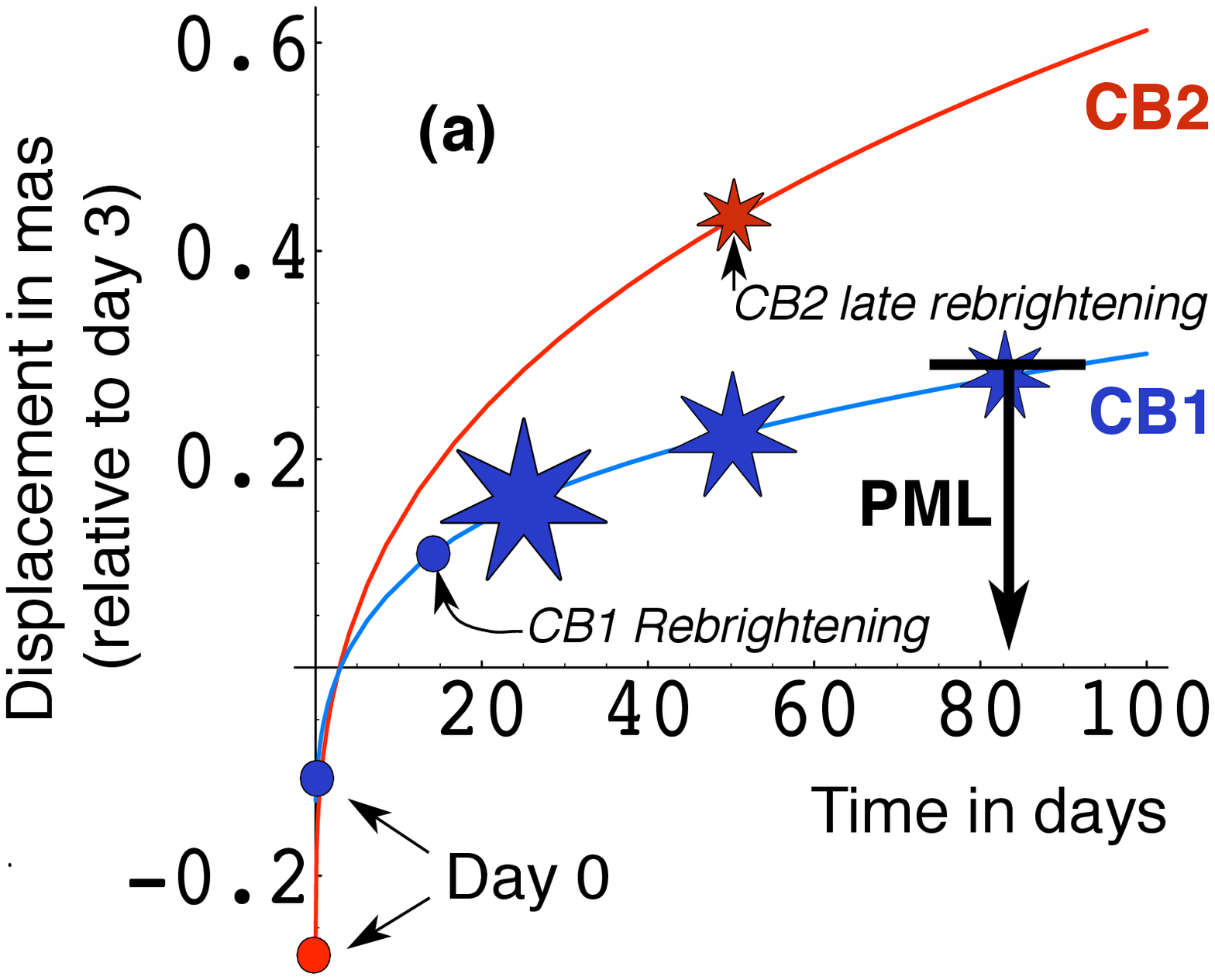, width=10cm}
\\
\hspace {1.7cm}
\epsfig{file=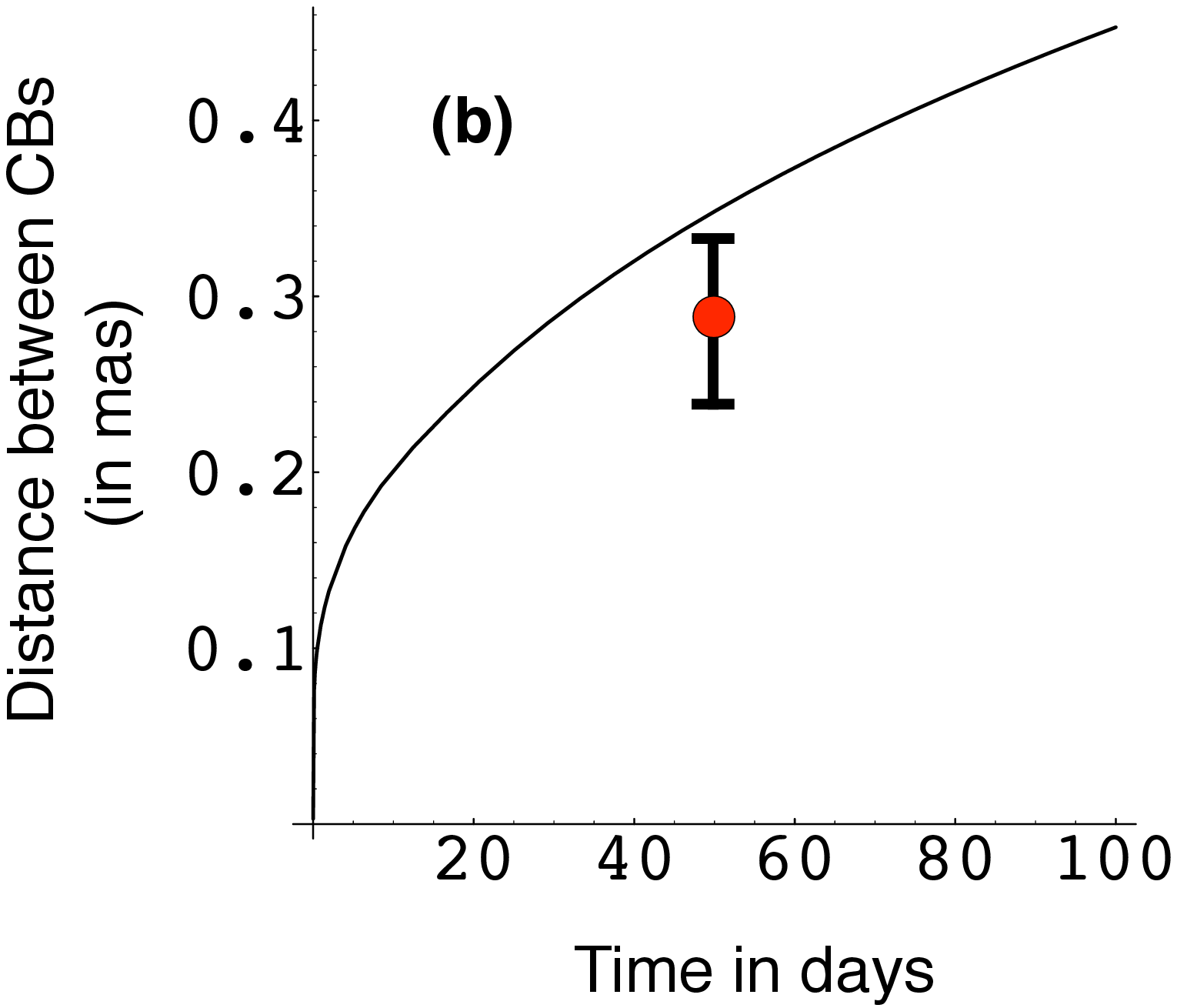, width=10cm}
\end{tabular}
\caption{{\bf (a)} The predicted angular displacement 
in the sky (in mas)  of the two CBs of GRB 030329,
as a function of observer's time from the 
first day of radio observations, day $\sim 3$. The positions at day 0, the
start-up time of the successive predicted rebrightenings of the slower CB1, 
 the observed time of the intense late
rebrightening of the faster CB2, as well as the fluences at 15.3 GHz on day 51
(70\% and 30\% of the total) are illustrated.
The proper motion limit (PML) of TFBK is also
shown, and discussed in Section 5.
{\bf (b)} The predicted angular distance between the two
CBs as a function of time, and its TFBK measurement at day 51, Eq.~(\ref{obs2}).}
\label{tororo}
\end{figure}

The (elaborate) results for the predicted motion of the two CBs,
relative to the first day of TFBK observations,
are shown in Fig.~(\ref{tororo}a).
The corresponding angular distance between the two CBs
as a function of time
is shown in Fig.~(\ref{tororo}b), along with the distance between
the two sources, measured by TFBK on day 51.

The central value $\Delta_{1,2}=0.28$ mas of Eq.~(\ref{obs2}) corresponds to
a transverse distance of $d=0.8$ pc at the GRB's location.
From Fig.~(\ref{tororo}a) one can read that, by
day 51, the faster (slower) CB should have moved $0.70$ mas (0.35 mas)
away from the parent SN, corresponding to a transverse distance
$d[2]=2.0$ pc ($d[1]=1.0$ pc). The distances along the CB's
trajectories are $d[2]/\theta[2]\sim 0.87$ kpc and $d[1]/\theta[1]\sim 0.43$ kpc.
The prediction of where a CB
is ---after travelling for hundreds of parsecs--- is not trivial,
particularly if the interstellar medium has the complicated density
profile required in the CB model to explain the intricate
optical AG light curves. A result that is correct to better than
$2\sigma$, as in Fig.~(\ref{tororo}b), is satisfactory.

\section{CB-model interpretation of the rest of the data}

The CBs emitted by microquasars are occasionally seen to ``rebrighten",
e.g. the western CB ejected, two years earlier, by XTE 1550-564 
(Corbel et al.~2002). We have interpreted in DDD the measured 
observer's time, $t\sim 1.5$ d ---when 
the optical light curves show a first fast rebrightening--- as the time at which
the faster CB reaches the stratified density profile at the boundary of the supperbubble.
The calculated time at which the slower CB reaches the same position is
$t\sim 13$ d. At that time the optical light curves are dominated by the
SN contribution and the second CB's rebrightening is barely observable.
Neither should the rebrightening be directly observable as a 
{\bf sharp} effect in the radio light-curves:  the radio emission
is delayed and smoothed by the time it takes electrons to ``cool down" to
radio-emitting frequencies (Dado et al.~2003a, DDD). For these reasons, and
because we have not developed the extremely laborious CB-model analysis of 
radio AGs in complicated density profiles, we did not pay attention in
DDD to the putative consequences of radio rebrightenings of the
CBs, neither did we report the predictions for the motion of the slower
CB and the distances between CBs, as we have now done in Fig.~(\ref{tororo}).
With the benefit of hindsight, {\bf these were oversights.} 

A very relevant rebrightening is that seen by Matheson et al.~(2003)
in the R-band AG of GRB 
030329, in observations beginning on day 51.75.
They report a ``jitter episode": {\it ``Variations of $>$ 30\% on timescales
of $\sim$ 2 days more than 50 days after  burst ... unlikely to
be in the SN component, as such variations have never been observed in
any other SN"}. The jitter is expected if one of the CBs is 
crossing new density inhomogeneities from day $\sim 50$ onwards. 
The rebrightening must be
very intense, since at that time the SN would otherwise be expected to be
very dominant, see Fig.~(\ref{figone}), and Fig.~(15) of Matheson et al.~(2003).

To summarize: The faster CB 
crosses various density 
variations (mainly enhancements) on days 1.5 to 7. The slower CB
is predicted to cross these enhancements and repeatedly rebrighten
from days $13$ to $\sim 60$.
The faster CB reaches new enhancements starting at day $\sim 50$.
Enhancements lead to rebrightenings, but the general trend
of a fading fluence steepens after a rebrightening: 
the fluence is proportional to a high power of the CB's Lorentz factor,
which diminishes fast while crossing the density enhancements  (DDD).
 { To accommodate the double source observed by TFBK, it must be that the 
faster CB has rebrightened, by day 51, to 30\% of the total radio signal at 
15.3 GHz, fading fast thereafter. This radio rebrightening and fading are expected, 
given the large optical ``jitter" observed at very similar times.}

How do we reconcile the observation of a double source at a single
frequency and a single time with the rest of the observations?

\begin{itemize}
\item{}
At day 51 the double source was observed at 15.4 GHz but not at
22.2 GHz. This may happen if the sensitivities differ, or if the CBs
have different spectra, as they should (they have different parameters
and they are crossing inhomogeneities at a given distance
at different observer's times and slightly different angles).
\item{} The predicted angular distances between the CBs at days
3 and 8 are unresolvably small, particularly if they are corrected
by the factor $(0.28\pm 0.5)/0.35$ discrepancy between the  
observation and the prediction, see Fig.~(\ref{tororo}b).
\item{}
Similarly corrected, the
predicted inter-CB angles at days 25 and 83 are $\sim 0.22\pm 0.04$ 
mas and $\sim 0.34\pm  0.06$ mas, which should have been 
resolvable. The slower CB must strongly dominate the fluence at
these dates. This is perfectly compatible with the CBs' rebrightening history,
summarized in Fig.~(\ref{tororo}a): at day 25 the slower CB has
recently rebrightened, and it is the only one observed. At day
51 the faster CB is strongly rebrightening, it is the extra source. 
After a very intense rebrightening the fluence decreases fast,
and by day 83 the faster CB has faded out of sight.
\end{itemize}

Two other items must be explained: the proper-motion limits of
Eq.~(\ref{obs3}), to be discussed in  \S5, and the  ``source sizes''
cited in \S2.
The sizes of astronomical ejecta may appear to be very different at
different wavelengths. An example is the radio galaxy Pictor A.
Observed in X-rays by Chandra, it shows a non-expanding jet that emanates 
from the centre of the galaxy and extends across $\sim 110$ kpc
 towards a brilliant hot spot $\sim 250$ kpc away  
(Wilson, Young \& Shopbell 2001). Observed at
1.4 GHz with VLA (Grandi et al. 2003) the diametrically opposed
jets have a somewhat biconical shape, with a large lateral extension.
In Dar \& De R\'ujula (2003a) we have argued that the extensive radio image
is due to  electrons that the CBs of Pictor A have bounced
in their voyage and ``non-collisionally" accelerated
to ``cosmic-ray" energies. These electrons may have
high energies, but their synchrotron radiation is only visible in
radio, given the low value of the intergalactic magnetic fields.
Contrariwise, synchrotron radiation by electrons within a CB's much
larger field produces photons of much higher energy, originating
from a very much more localized source.

Observations such as the above one imply  that the CBs of
GRBs, though  effectively ``point-like'' in their visible or X-ray
emission,  may be ``sizeable" in the radio. The sizes observed
by TFBK may be the sizes of the CB-induced forward cone of electrons,
not of the CBs themselves. These sizes
may have a complicated time dependence, since they are functions
of the ambient densities and magnetic fields that the CBs encounter. 

\section{Proper motion limits on the CB model}

TFBK state:
{\it 
``Dar \& De R\'ujula 
(2003b) predicted a displacement of 2 mas over the 80 days
of our VLBI experiment assuming plasmoids propagating in a constant
density medium. This estimate was revised downward to 0.55 mas by
incorporating plasmoid interactions with density inhomogeneities at a
distance of $\sim$100 pc within a wind-blown medium 
(Dado, Dar \& De R\'ujula 2004). Neither variant of this model is consistent 
with our proper motion limits.''}

TFBK are right in stating that the DDD prediction for the motion
of a {\it single} source is not what is observed\footnote{The proper-motion
 limit in Eq.~(\ref{obs3}) is, strictly
speaking, not a test of the proper motion predicted by the 
CB model. This is because TFBK presumably tested the hypothesis
of a displacement
$\vec{d}= \vec v\,t$, with $\vec v$ a constant. In the CB model,
CBs decelerate and the prediction is, as in Eq.~(\ref{vcbang}),
$\vec{d}\approx\vec w\;t^{1/3}$ with $\vec w$ a constant vector (DDD).
Given the relatively large error bars, this point may be minor.},
though it is difficult to rule out a proper motion
with confidence with data
containing 20$\sigma$ evidence for two sources separated by 28 mas.
This datum and the
proper-motion limits are inconsistent at first sight. 
If the radio luminosity of the slower CB dominates
the radio AG after its rebrightening, there is no contradiction with
the proper motion limit, as shown in Fig.~(\ref{tororo}a). The limit refers 
to a single source, the central value of the circular-Gaussian 
fits to the data. In our interpretation, that source is the slow CB.

\section{Scintillations}

TFBK state:
{\it 
``Strong and persistent intensity variations in centimeter radio 
light curves for}
{\bf all} {\it
GRBs are expected in the cannonball model. Strong intensity variations
are not seen for GRB 030329 ...  There
are moderate variations seen in the radio light curves of GRB 030329 
(25\%
at 4.9 GHz, 15\% at 8.5 GHz and 8\% at 15 GHz) which decrease by a
factor of three from $\sim$3 to 40 d after the burst."}
That {\it ``strong intensity variations are not seen''} is, we believe,
an overstatement. The radio light curves (Sheth et al. 2003; Berger et al. 2003) 
show intensity variations that
at 4.86 GHz are close to a factor of 2 through day 10.

The observed trend of intensity variations diminishing
with time can be understood within the CB model (Dado et al.~2003a, DDD).
They are very reminiscent of 
the ones seen in radio signals from pulsars  in  the Galaxy.
Gupta (1995) demonstrated for a sample of 59 pulsars that their 
transverse speed, $V_{iss}$, measured from their interstellar 
scintillations, agrees well with the value, $V_{pm}$,  
directly measured as proper motion (see also Nicastro et al.~2001).  
The mean $V_{pm}$ of Gupta's pulsars is 311 $\rm km\, s^{-1}$ 
and their mean distance $\langle D\rangle$ is  $\sim$ 1.96 kpc. 
Their angular speeds are within an order of magnitude of a central value
$\dot\alpha_{pm}={\langle V_{pm}\rangle / \langle D\rangle} 
\simeq 5.1\times 10^{-15}$ rad s$^{-1}$. Such
angular velocities 
result in observed scintillations {\it ``with a modulation index of order unity and a
time scale of a few hours"} (TFBK).

The predicted angular velocities of the CBs can be estimated 
with use of  Eq.~(\ref{omega}) and the effective
$x_\infty$ values quoted in Section 3. They are 
$\dot\alpha_{_{\rm CB}}\sim 1.7 \times 10^{-15}$ rad s$^{-1}$
and $\sim 8.6 \times 10^{-16}$ rad s$^{-1}$ for the faster and slower
CB at day 3. These are somewhat smaller than
the typical pulsar values and should result in
the observed {\it ``modulation index of order unity and time scale of a few 
hours"} at the smaller VLBI radio frequencies.
On day 40, the other date chosen by TFBK in discussing scintillations,
the predictions are 
$\dot\alpha_{_{\rm CB}}\sim 2.9 \times 10^{-16}$ rad s$^{-1}$
and $\sim 1.4 \times 10^{-16}$ rad s$^{-1}$ for the faster and slower (dominant)
CB. These values are more than one order of magnitude smaller
than the typical pulsar values. By then, the expected scintillations
should be similar to those observed in very slowly moving pulsars,
which are mainly due to the motions of the Earth, the Sun and the
turbulent interstellar medium. Their modulation index is far below
unity and their time scale is days long. This evolution
towards less pronounced scintillations is precisely the trend
of the GRB 030329 data. For this reason, we do not agree 
with TFBK that the absence
of rapid fluctuations is a problem for the CB model. 

We gave an explanation in Dar \& De R\'ujula (2003a) why the
observed jets of CBs are
wider in radio than at higher frequencies, and extracted here 
the consequence that the CBs of GRBs may have observable radio sizes (\S 4). 
A size increasing with time would progresively
quench scintillations, as in standard models (Frail et al.~1997). 

\section{Conclusions}

TFBK conclude: {\it ``Much less easy to explain is
the single observation 51 days after the burst of an additional
radio component 0.28 mas northeast of the main afterglow.  This
component requires a high average velocity of $19\, c$ and cannot
be readily explained by any of the standard models.  Since it is
only seen at a single frequency, it is remotely
possible that this image is an artifact of the calibration."}

We have interpreted the double source discovered by TFBK 
as an image of the two cannonballs required in the CB model to explain
the double-peaked shape of GRB 030329 and the double-shoulder
nature of its optical AG. 
The observed separation between the CBs is roughly the separation
expected from the CB-model fit in DDD to the optical AG of GRB 030329.
But the main point is that the two CBs  appear to have been observed,
and their separation is ``hyperluminal". Seen in this light, the double source
observed by TFBK is an extraordinarily important discovery in GRB physics,
rather than a 20$\sigma$ fly in the ointment.

\vskip 1cm
{\bf Acknowledgements.} We are grateful to Greg Taylor for
having sent us a poster containing a preliminary version of the
TFBK results, and for useful discussions on version 1 of
this paper. We are indebted to Andy Cohen and Shelly Glashow
for useful comments and suggestions. 
This research was supported in part by the Helen
Asher Space Research Fund and by the VPR fund for research
at the Technion.

\end{document}